\documentstyle[12pt]{article}
\makeatletter
\gdef\@pubnumber{\null}
\long\def\pubnumber#1{\gdef\@pubnumber{DAMTP 93-46}}
\def\@makepub{\vbox to \z@{\hbox to
\textwidth{\hfill\llap{\parbox[t]{0.33\textwidth}{\raggedleft\@pubnumber}}}%
\vss}}
\def\@maketitle{\newpage
\@makepub \null
\vskip 2em \begin{center}
{\LARGE \@title \par} \vskip 1.5em {\large \lineskip 0.5em
\begin{tabular}[t]{c}\@author
\end{tabular}\par}
\vskip 1em {\large \@date} \end{center}
\par
\vskip 1.5em}

\pubnumber{1}

\begin{document}

\title{Charged Particle Scattering From Electroweak and Semi-local Strings}
\author{A.C.Davis\thanks{and King's College, Cambridge} and
	A.P.Martin,\\
	{\normalsize DAMTP,}\\
	{\normalsize{Cambridge University,}}\\
	{\normalsize{Silver St.,}}\\
	{\normalsize{Cambridge}}\\
	{\normalsize{CB3 9EW}}\\
	{\normalsize{U.K.}}\\
	\\
	N.Ganoulis,\\
	{\normalsize{Physics Division,}}\\
	{\normalsize{School of Technology,}}\\
	{\normalsize{Thessaloniki,}}\\
	{\normalsize{Greece.}}}

\maketitle

\begin{abstract}

The scattering of a charged fermion from an electroweak or semi-local
string is investigated and a full solution obtained for both massive
and massless cases. For the former, with fractional
string flux, there is a helicity conserving and helicity-flip
cross-section, of equal magnitude and of a modified Aharonov-Bohm
form: for integer flux the strong interaction
cross-section is suppressed by a logarithmic term. The results also
apply for GUT cosmic strings and chiral fermions.

\end{abstract}

\pagebreak

\section{Introduction}

The existence of string defects in models displaying spontaneous
symmetry-breaking with topologically non-trivial vacuum manifolds has
been known for some time \cite{vil}. Many grand unified
theories, for example, possess local string solutions
and the axion model of Peccei
and Quinn \cite{pq} has a global string solution. The simplest model of
a local string is the Nielsen-Olesen vortex \cite{no}. For fractional
string flux the dominant scattering for a charged particle off such a
string is Aharonov-Bohm \cite{alfw}\cite{wbp2}, whilst if the
string flux is
integer, one obtains the scattering cross-section of Everett \cite{eve}.
Recently, a wider class of string defects, the semi-local
\cite{av}\cite{hind}
and the electroweak strings \cite{vach} have been discovered in theories where
the first homotopy group of the vacuum manifold is topologically
trivial.
These, the embedded
defects, are at best metastable. Since this new class of string
defects has special features, it is interesting to see if they also
exhibit Aharonov-Bohm scattering. This is the aim of this paper.

The Aharonov-Bohm scattering of fermions
off electroweak strings has been studied
by one of us before \cite{gan}. There it was found that the scattering
cross-section violated helicity, and went to zero in the limit of
zero mass particles. However, due to the complexity of the problem a
full solution was not obtained and the dominant mode approximation
used. In this paper we are able to go beyond this approximation to
obtain a fuller scattering solution using the `top-hat' model of the
string-core, discussed in \cite{wbp2}. There it was shown that the
scattering cross-section was insensitive to the core-model, and the
simple `top-hat' model gave the same result as the more sophisticated
one. Using this method we are able to clarify the results of \cite{gan}
and show that, in the massive case, for fractional flux there are both
helicity conserving and helicity-flip
Aharonov-Bohm scattering cross-sections with the
electroweak string. The helicity violating cross-section is seen to go
to zero in the limit of zero mass.
We also show that there is an analogue of the
Everett cross-section in the case of integer flux and massive
particles..
 This is also the
case for semi-local strings. Hence, for both electroweak and semi-local
strings with fractional flux the elastic-scattering cross-section is a
strong interaction
cross-section, independent of the string radius. For integer flux
there is only a logarithmic suppression of the strong interaction
cross-section. Thus, if such strings are metastable they will interact
with the surrounding plasma in an analogous way to local topological
strings \cite{kibb}.

The plan of this paper is as follows:- In section 2 we review the
semi-local and electroweak strings. We also recap the work of \cite{gan}
for scattering from electroweak strings. In section 3 we calculate the
scattering amplitudes of fermions in interaction with
electroweak strings in both massive and massless cases, making use
of the `top-hat' core model and in section 4 we obtain
the corresponding cross-sections in the cases of integer and
non-integer flux. This method is also applied to the case of fermions
scattering off semi-local strings in section 5, our conclusions on
this work being contained in section 6.

\newpage

\section{Semi-local and Electroweak Strings}

The general requirement
for the formation of topological string defects is that the vacuum manifold is
multiply connected. The simplest model of this is the U(1)
Nielsen-Olesen vortex \cite{no}. Recently, however, Vachaspati and
Ach\'{u}carro \cite{av} demonstrated that if
{\em{both}} global {\em{and}} local symmetries are present in a model then
it is possible to have stable string solutions even if the vacuum
manifold is simply connected. The fact that they share properties
with both local and global defects led to them being dubbed semi-local strings.

The string investigated in \cite{av} arises in an
extension of the Abelian Higgs model where the complex scalar field is
replaced by an SU(2) doublet, $\Phi$, such that the action is invariant under
G=SU(2)$_{g}\times$U(1)$_{l}$ where $g$ and $l$ denote global and local
symmetries respectively. When $\Phi$ condenses, breaking G to
H=U(1)$_{l}$  the corresponding vacuum manifold is the 3-sphere,
S$^{3}$. This is clearly simply connected and
the formation of string defects would not normally be expected.
If, however, we now consider the gauge
component of the symmetries alone, then
it is seen that the vacuum manifold is a circle, and, hence, not
simply connected. This suggests the existence of string solutions.
The way to reconcile this apparent paradox is to
consider the change in energy in passing between such solutions.

For each point on the vacuum manifold, the U(1)$_{l}$ transformation
generates a circle around the three-sphere; each of these circular
paths, in turn,
corresponds to a string solution. Hence, we can think of the vacuum
manifold as comprising of the set of all possible string
solutions. To move to another point on the manifold along one of these
circular paths
involves a gauge transformation and so expends no energy. However, to
leave such a path involves a global transformation and thus requires a
change in energy. These energy barriers are what enable string
solutions to exist on a topologically trivial manifold, since to
contract a loop to a point (the normal way of removing such defects)
always involves a cost in energy. It has been shown that the stability
of semi-local strings is parameter dependent, depending on the ratio
of the masses of the Higgs and vector particles \cite{hind}.

Now, the extended Abelian-Higgs model
that yields the semi-local string is none other than the
Electroweak model with zero SU(2) charge. This led to the postulation
of non-topological, yet metastable, strings in the Electroweak
theory \cite{vach}. Indeed, such a solution is found,
and bears a predictable
likeness to the semi-local case.The existence
of such strings is closely linked to that of the semi-local
string. Whilst the string solution is unstable in the minimal
electroweak theory \cite{marg}, extensions may possess stable string
solutions.

Consider the string solution in the standard electroweak theory with
fermionic interactions, the relevant part of the Lagrangian
being
\begin{equation}
	{\cal{L}}=
	i\bar{L}\gamma^{\mu}D_{\mu}L+
	i\bar{e}_{R}\gamma^{\mu}D_{\mu}e_{R}-
	f_{e}(
	\bar{L}e_{R}\Phi+\Phi^{\dagger}{\bar{e}}_{R}L)
\end{equation}
where $\bar{L}=(\bar{\nu},{\bar{e}}_{L})$, $f_{e}$ is the Yukawa
coupling, $\Phi$ is the usual Higgs field and the
covariant derivative is given by
\begin{equation}
	D_{\mu}=\partial_{\mu}+\frac{i\alpha\gamma}{2}Z_{\mu}
\end{equation}
where $\gamma=e/(\sin{(\theta_{W})}\cos{(\theta_{W})})$ and $\theta_{W}$ is the
Weinberg angle. The $Z$-coupling, $\alpha$, has the form
\begin{equation}
	\alpha=-2(T_{3}-Q\sin^{2}{\theta_{W}})
\end{equation}
where $T_{3}$ is weak isospin and $Q$ is electric charge. This clearly
varies according to the field to which it is coupled. In particular we
note that for electrons and down quarks,
\begin{equation}
	\alpha_{L}=\alpha_{R}+1
\end{equation}
so we have a marked asymmetry between left and right
fields\footnote{For up quarks the relationship is reversed with
$\alpha_{L}=\alpha_{R}-1$, but we still have an asymmetry between left
and right fields}.

In \cite{gan}, when considering the scattering of fermions off an
electroweak string,
use was made of the solution derived in \cite{vach}, where a U(1)
Nielsen-Olesen string is embedded in SU(2)$\times$U(1). More explicitly
\begin{equation}
	\begin{array}{lcl}
	\Phi = \left(
	\begin{array}{c}
	\Phi^{+}\\
	\Phi^{0}\\
	\end{array}
	\right) &
	= & f(r)e^{i\phi}\left(
	\begin{array}{c}
	0\\
	1\\
	\end{array}
	\right),\\
	 & & \\
	Z_{\phi} = -\frac{v(r)}{r}, &  & Z_{r} = 0, \\
	 & & \\
	W = 0, &  & A = 0,\\
	\end{array}
\end{equation}
where $Z$ and $W$ are the usual gauge bosons, and $A$ is the photon
field. The functions $f$ and $v$ are functions of $r$ only: they are
the usual Nielsen-Olesen solutions, found by solving the appropriate field
equations and have approximate profiles
\begin{eqnarray}
	f(r) & = &
	\cases{\frac{\eta}{\sqrt{2}}&$r\geq R$\cr
	\frac{\eta}{\sqrt{2}}\left(\frac{r}{R}\right)&$r<R$\cr}\nonumber\\
	v(r) & = &
	\cases{\frac{2}{\gamma}&$r\geq R$\cr
	\frac{2}{\gamma}\left(\frac{r}{R}\right)^{2}&$r<R$\cr}\nonumber\\
	\nonumber
\end{eqnarray}
We note that since $Z$ and $\Phi$ are independent of $z$ the equations
of motion will be, essentially, 2+1 dimensional Dirac equations.

Considering the case of electrons only for the time being, if we set
$m=f_{e}\eta/\sqrt{2}$ and $\partial_{t}=-i\omega$ where $\omega$ is the
energy of the electron, then taking the usual Dirac representation and
writing $e_{L}=(0,\psi)$, $e_{R}=(\chi,0)$ we obtain the following
equations of motion for $\psi$ and $\chi$;
\begin{equation}
	\begin{array}{rcrcrcr}
	\omega\chi & + & i\sigma^{j}D^{R}_{j}\chi & - &
	f_{e}fe^{-i\phi}\psi & = & 0,\\
	\omega\psi & - & i\sigma^{j}D^{L}_{j}\psi & - &
	f_{e}fe^{i\phi}\chi & = & 0.\\
	\end{array}
	\label{ganeq}
\end{equation}
Note the phases $e^{\pm i\phi}$ and the coupling of $\psi$ to $\chi$ via the
mass term. The effects of these, plus the different couplings to the
$Z$, manifest themselves in a non-zero amplitude for helicity
flip.

To see that helicity is not conserved consider the following.
{}From (\ref{ganeq}) it is easy to write down the Hamiltonian for the
system using canonical momenta;
$\underline{\sigma}.\underline{\pi}=-i\sigma^{j}D_{j}$
\begin{equation}
	H=
	\left(
	\begin{array}{cc}
	-\underline{\sigma}.{\underline{\pi}}_{R} &
	f_{e}fe^{-i\phi}\\
	f_{e}fe^{i\phi} &
	\underline{\sigma}.{\underline{\pi}}_{L}\\
	\end{array}
	\right).
\end{equation}
The helicity
operator is defined as
\begin{equation}
	\underline{\Sigma}.\underline{\Pi}=
	\left(
	\begin{array}{cc}
	\underline{\sigma}.{\underline{\pi}}_{R} &
	0\\
	0 &
	\underline{\sigma}.{\underline{\pi}}_{L}\\
	\end{array}
	\right).
\end{equation}
If helicity was conserved one would expect the commutator of the
Hamiltonian with the helicity operator to be zero. However, when
calculated, one finds that
\begin{equation}
	[H,\underline{\Sigma}.\underline{\Pi}]=if_{e}
	\left(
	\begin{array}{cc}
	0 &
	\sigma^{j}(D^{j}\phi^{0})^{*}\\
	\sigma^{j}D^{j}\phi^{0} &
	0\\
	\end{array}
	\right),
\end{equation}
which is non-zero inside the string core. This non-conservation of
helicity differs from the usual Aharonov-Bohm scattering off a
thin solenoid of magnetic flux only. It is noted, however, that if we
take the massless case where $f_{e}=0$ and we have no coupling
betwen the $\psi$ and $\chi$, then
$[H,\underline{\Sigma}.\underline{\Pi}]=0$ and helicity will
be conserved. Hence, we expect the cross-sections for the massless and
massive cases to differ.

Considering the case of an incoming plane wave of positive helicity,
it can be demonstrated \cite{gan} that one mode tends to dominate the
scattering,
the cross-sections for this mode being identical for positive and
negative helicity scattered states, such that, to leading order,
\begin{equation}
	\left.\frac{d\sigma}{d\theta}\right|_{\pm}\sim
	\frac{1}{k}\left(
	\frac{\omega-k}{2\omega}\right)^{2}
	\sin^{2}{(\pi\alpha_{R})}
\end{equation}
where $k$ is the momentum of the electron. This is, however, only a
partial result, so a natural next step is to try and obtain a fuller
result by means of a simpler profile for $Z$ and $\Phi$.

\newpage

\section{Scattering from the Electroweak String}

Our method follows those of \cite{gan} and \cite{wbp2}, though we
adopt the simpler ``top-hat''
model in an attempt to
get a fuller result:
\begin{equation}
	f(r)=\left\{
	\begin{array}{lll}
 		0 & , & r<R \\
		\eta/\sqrt{2} & , & r>R \\
	\end{array}
	\right.
\end{equation}
\begin{equation}
	v(r)=\left\{
	\begin{array}{lll}
 		0 & , & r<R \\
		2/\gamma & , & r>R \\
	\end{array}
	\right.
\end{equation}
The equations of motion are the same as in (\ref{ganeq}), up to the
altered gauge fields and covariant derivatives.
We note that although the $\chi$ and $\psi$ are coupled via
the mass term, using the above profile, there is no mass inside the
core and
hence no coupling.

We try the usual mode decomposition

\begin{equation}
	\chi(r,\phi) =  \sum_{l=-\infty}^{\infty}
	\left(
	\begin{array}{c}
		\chi_{1}^{l}(r)\\
		i\chi_{2}^{l}(r)e^{i\phi}\\
	\end{array}
	\right)
	e^{il\phi},
	\hspace{.5cm}
	\psi(r,\phi)  =  \sum_{l=-\infty}^{\infty}
	\left(
	\begin{array}{c}
		\psi_{1}^{l}(r)\\
		i\psi_{2}^{l}(r)e^{i\phi}\\
	\end{array}
	\right)
	e^{i(l+1)\phi}.\\
	\label{dec1}
\end{equation}
Making use of

\begin{equation}
	\sigma^{j}D_{j}=
	\left(
	\begin{array}{cc}
		0 & e^{-i\phi}(D_{r}-iD_{\phi}) \\
		e^{i\phi}(D_{r}+iD_{\phi}) & 0 \\
	\end{array}
	\right)
	\label{sig}
\end{equation}
\vspace{.5cm}
we now substitute (\ref{dec1}) and (\ref{sig}) into
(\ref{ganeq}) to obtain

\begin{equation}
	\begin{array}{rcrcrcl}
	\omega\chi^{l}_{2} & + &

	(\frac{d}{dr}-\frac{l}{r}+\frac{\alpha_{R}\gamma v}{2r})\chi^{l}_{1}
	& - &
	f_{e}f\psi_{2}^{l} & = & 0, \\
	 & & & & & & \\
	\omega\chi^{l}_{1} & - &
	(\frac{d}{dr}+\frac{l+1}{r}-\frac{\alpha_{R}\gamma v}{2r})\chi^{l}_{2}
	& - &
	f_{e}f\psi_{1}^{l} & = & 0, \\
	 & & & & & & \\
	\omega\psi^{l}_{2} & - &
	(\frac{d}{dr}-\frac{l+1}{r}+\frac{\alpha_{L}\gamma v}{2r})\psi^{l}_{1}
	& - &
	f_{e}f\chi_{2}^{l} & = & 0, \\
	 & & & & & & \\
	\omega\psi^{l}_{1} & + &
	(\frac{d}{dr}+\frac{l+2}{r}-\frac{\alpha_{L}\gamma v}{2r})\psi^{l}_{2}
	& - &
	f_{e}f\chi_{1}^{l} & = & 0. \\
	\end{array}
	\label{eqs}
\end{equation}
We also insist that our solutions are eigenfunctions of the
helicity operator,
which implies that
\begin{equation}
	\begin{array}{rcr}
	\underline{\sigma}.\underline{\pi}_{R}\chi & = & \pm k \chi,\\
	\underline{\sigma}.\underline{\pi}_{L}\psi & = & \pm k \psi,\\
	\end{array}
\end{equation}
where $k$ is the momentum, and $+$($-$) corresponds to
positive(negative) helicity. On substitution in (\ref{eqs}) this yields
\begin{equation}
	\begin{array}{rcr}
	(\omega\mp k)\chi & = & f_{e}f\psi, \\
	(\omega\pm k)\psi & = & f_{e}f\chi, \\
	\end{array}
	\label{hel}
\end{equation}
so giving us a relation between $\chi$ and $\psi$.

We now need to consider solutions of the Dirac equation inside and outside the
string core.

\subsection*{(a) Internal solution: $r<R$}
Taking the profiles described earlier we have $f=v=0$ for $r<R$,
so our equations of motion
(\ref{eqs}) reduce to
\begin{equation}
	\begin{array}{rcrcl}
	\omega\chi^{l}_{2} & + &
	(\frac{d}{dr}-\frac{l}{r})\chi^{l}_{1}
	& = & 0,\\
	 & & & & \\
	\omega\chi^{l}_{1} & - &
	(\frac{d}{dr}+\frac{l+1}{r})\chi^{l}_{2}
	&  = & 0, \\
	 & & & & \\
	\omega\psi^{l}_{2} & - &
	(\frac{d}{dr}-\frac{l+1}{r})\psi^{l}_{1}
	& = & 0, \\

	 & & & & \\
	\omega\psi^{l}_{1} & + &
	(\frac{d}{dr}+\frac{l+2}{r})\psi^{l}_{2}
	& = & 0. \\
	\end{array}
	\label{eqin}
\end{equation}
Consider first the $\chi$. Combining the top two equations of (\ref{eqin})
and setting $z=\omega r$ we obtain

\begin{equation}
	\begin{array}{rcrcl}
	\frac{1}{z}\frac{d}{dz}(z\frac{d}{dz})\chi_{1}^{l} & + &
	(\frac{z^{2}-l^{2}}{z^{2}})\chi_{1}^{l} & = & 0,\\
	\end{array}
\end{equation}
which is easily recognised as Bessels equation of order $l$. Hence, by
square integrability and regularity at the origin, the internal solution is

\begin{equation}
	\chi_{1}^{l}=c_{l}J_{l}(\omega r).
\end{equation}
Now, $\chi_{1}^{l}$ and $\chi_{2}^{l}$ are coupled via (\ref{eqin})
so,
making use of the Bessel function
relations\footnote{These actually hold
for Bessel functions of the first, second or
third kind.}
\begin{equation}
	\begin{array}{ccccccc}
	\frac{d}{dz}J_{\mu}(z) & = &
	\frac{1}{2}(J_{\mu-1}(z)-J_{\mu+1}(z)) & , &
	\frac{\mu}{z}J_{\mu}(z) & = &
	\frac{1}{2}(J_{\mu-1}(z)+J_{\mu+1}(z)),
	\end{array}
	\label{bess}
\end{equation}
it is easy to show that
$\chi_{2}^{l}  =  c_{l}J_{l+1}(\omega r)$.

Inside the core, however, $f=0$ so, unfortunately,
(\ref{hel}) gives us no information
about a link between $\chi$ and $\psi$. From (\ref{eqin}), though, it
is easy to see that $\psi_{1}^{l}$ and $\psi_{2}^{l}$ will satisfy
Bessel equations of order $(l+1)$ and $(l+2)$ respectively, so our
internal solution is
\begin{equation}
	\left(
	\begin{array}{c}
	\chi\\
	\psi\\
	\end{array}
	\right)
	=
	\sum_{l=-\infty}^{\infty}
	\left(
	\begin{array}{c}
	c_{l}J_{l}(\omega r)\\
	ic_{l}J_{l+1}(\omega r)e^{i\phi}\\
	d_{l}J_{l+1}(\omega r)e^{i\phi}\\
	id_{l}J_{l+2}(\omega r)e^{2i\phi}\\
	\end{array}
	\right)
	e^{il\phi}
\end{equation}
where $c_{l}$ and $d_{l}$ are independent.

\newpage

\subsection*{(b) External solution: $r>R$}

For large r, we now take $f(r)=\frac{\eta}{\sqrt{2}}$ and
$\frac{v(r)}{r}=\frac{2}{\gamma r}$,
so that, defining
$m=\frac{f_{e}\eta}{\sqrt{2}}$,
our equations of motion become

\begin{equation}
	\begin{array}{rcrcrcl}
	(\frac{d}{dr}-\frac{l}{r}+\frac{\alpha_{R}\gamma v}{2r})\chi^{l}_{1}
	& + &\omega\chi^{l}_{2} &- &
	m\psi_{2}^{l} & = & 0, \\
	 & & & & & & \\
	(\frac{d}{dr}+\frac{l+1}{r}-\frac{\alpha_{R}\gamma v}{2r})\chi^{l}_{2}
	& - & \omega\chi^{l}_{1} & + &
	m\psi_{1}^{l} & = & 0, \\
	 & & & & & & \\
	(\frac{d}{dr}-\frac{l+1}{r}+\frac{\alpha_{L}\gamma v}{2r})\psi^{l}_{1}
	& - & \omega\psi^{l}_{2} & + &
	m\chi_{2}^{l} & = & 0, \\
	 & & & & & & \\
	(\frac{d}{dr}+\frac{l+2}{r}-\frac{\alpha_{L}\gamma v}{2r})\psi^{l}_{2}
	& + & \omega\psi^{l}_{1} & - &
	m\chi_{1}^{l} & = & 0. \\
	\end{array}
	\label{eqout}
\end{equation}
Setting

\begin{displaymath}
	\begin{array}{ll}
	L=	\left(
		\begin{array}{cccc}
		-l & 0 & 0 & 0\\
		0 & l+1 & 0 & 0\\
		0 & 0 & -l-1 & 0\\
		0 & 0 & 0 & l+2\\
		\end{array}
		\right), &
	C=	\left(
		\begin{array}{cccc}
		0 & 1 & 0 & 0\\
		-1 & 0 & 0 & 0\\
		0 & 0 & 0 & -1\\
		0 & 0 & 1 & 0\\
		\end{array}
		\right),\\
	 & \\
	Q=	\left(		\begin{array}{cccc}
		\alpha_{R} & 0 & 0 & 0\\
		0 & -\alpha_{R} & 0 & 0\\
		0 & 0 & \alpha_{L} & 0\\
		0 & 0 & 0 & -\alpha_{L}\\
		\end{array}
		\right), &
	P=	\left(
		\begin{array}{cccc}
		0 & 0 & 0 & -1\\
		0 & 0 & 1 & 0\\
		0 & 1 & 0 & 0\\
		-1 & 0 & 0 & 0\\
		\end{array}
		\right).\\
	 & \\
	\end{array}
\end{displaymath}
and $w_{l}=(\chi_{1}^{l},\chi_{2}^{l},\psi_{1}^{l},\psi_{2}^{l})$
we can write (\ref{eqout}) as

\begin{equation}
	\frac{d}{dr}w_{l}  +  \frac{1}{r}(L+Q)w_{l}  +
	(\omega C + mP)w_{l}  =  0
	\label{eqout2}
\end{equation}
the asymptotic behaviour of which is decided by the eigenvalues of
$(\omega C + mP)$; found to be $\pm ik$ as one would expect.

Defining $N=L+Q$ and $\nu=l-\alpha_{R}$, we can use our relation
$\alpha_{L}-\alpha_{R}=1$
to write $N$ as
\begin{equation}
	N=\left(
	\begin{array}{cccc}
	-\nu & 0 & 0 & 0\\
	0 & \nu+1 & 0 & 0\\
	0 & 0 & -\nu & 0\\
	0 & 0 & 0 & \nu+1\\
	\end{array}
	\right).
\end{equation}
{}From this we see that (\ref{eqout2}) is the radial part of the free
Dirac equation
with shifted angular momentum. Taking the second derivative
of (\ref{eqout2}), it can be shown that this reduces to

\begin{equation}
	\frac{d^2}{dr^2}w_{l}+\frac{1}{r}\frac{d}{dr}w_{l}
	-\frac{N^2}{r^2}w_{l}+k^{2}w_{l}=0.
\end{equation}
Thus $\chi_{1}^{l},\chi_{2}^{l},\psi_{1}^{l},\psi_{2}^{l}$  obey
independent Bessels equations of differing orders. By
(\ref{hel}) however, for $r>R$,

\begin{equation}
	\begin{array}{rcr}
	(\omega\mp k)\chi & = & m\psi, \\
	(\omega\pm k)\psi & = & m\chi. \\
	\end{array}
\end{equation}
These together with (\ref{eqout}) imply that
\begin{equation}
	\chi_{2}^{l}=\mp \left( \frac{d}{dr}-\frac{\nu}{r}\right)\chi_{1}^{l}
\end{equation}
where the upper (lower) sign corresponds to a positive (negative)
helicity.
Making use of our Bessel function relations (\ref{bess}) once more,
we see, then, that the
general solution for $r>R$ is

\begin{equation}
	\left(
	\begin{array}{c}
	\chi\\
	\psi\\
	\end{array}
	\right)
	=
	\sum_{l=-\infty}^{\infty}
	\left(
	\begin{array}{c}
	Z_{\nu}(k r)\\
	\pm iZ_{(\nu+1)}(k r)e^{i\phi}\\
	B^{\pm}Z_{\nu}(k r)e^{i\phi}\\
	\pm iB^{\pm}Z_{(\nu+1)}(k r)e^{2i\phi}\\
	\end{array}
	\right)
	e^{il\phi}
\end{equation}
where $B^{\pm}=m/(w\pm k)$ and $Z_{\nu}$ is a Bessel function
of order $\nu$. It is important to realise that when substituting in
for the Bessel function, it is necessary to pay close attention to the
sign of the order: for order $\nu$ we will have
$Z_{(\nu+1)}=J_{\nu+1},N_{\nu+1}$ whilst for $-\nu$ we will have
$Z_{(\nu+1)}=-J_{-(\nu+1)},-N_{-(\nu+1)}$. Also note that we have a $k$
in the Bessel
function arguments, as opposed to the $\omega$ in the internal
solution. This arises because the mass
outside the core is non-zero, so, as $\omega^{2}=m^{2}+k^{2}$,
$\omega\neq k$.

\newpage

\subsection*{(c) Asymptotic solution}

We require, now, that our solution be an incoming planewave of, say,
positive
helicity, plus a scattered wave consisting of both positive {\em and}
negative helicity states, since we know that helicity is not conserved
in the case of finite mass.
To apply this we need to equate our general external solution (\ref{doog})
with such an asymptotic form;
\begin{displaymath}
	\sum_{l=-\infty}^{\infty}
	\left(
	\begin{array}{c}
	(-i)^{|l|}J_{|l|}e^{il\phi}\\
	\pm i(-i)^{|l|}J_{\pm(l+1)}e^{i(l+1)\phi}\\
	B^{+}(-i)^{|l|}J_{|l|}e^{i(l+1)\phi}\\
	\pm iB^{+}(-i)^{|l|}J_{\pm(l+1)}e^{i(l+2)\phi}\\
	\end{array}
	\right)+
	\frac{f_{l}e^{ikr}}{\sqrt{r}}
	\left(
	\begin{array}{c}
	e^{il\phi}\\
	e^{i(l+1)\phi}\\
	B^{+}e^{i(l+1)\phi}\\
	B^{+}e^{i(l+2)\phi}\\
	\end{array}
	\right)+
	\frac{g_{l}e^{ikr}}{\sqrt{r}}
	\left(
	\begin{array}{c}
	e^{il\phi}\\
	-e^{i(l+1)\phi}\\
	B^{-}e^{i(l+1)\phi}\\
	-B^{-}e^{i(l+2)\phi}\\
	\end{array}
	\right)
\end{displaymath}
with $+(-)$ for $l>$($<$)0.
Here the first term is an incoming (and outgoing) plane wave of
positive helicity, and
the second and third terms are the positive and negative helicity
components of the scattered wave respectively.

There are two cases to consider:

\paragraph{(i) \boldmath$\nu\leq$-1, 0\boldmath$\leq\nu$:}
We take as our two independent solutions $Z_{\nu}^{1}=J_{|\nu|}$ and
$Z_{\nu}^{2}=N_{|\nu|}$. Matching coefficients of $e^{il\phi}$ we obtain
\begin{displaymath}
	\begin{array}{rcrcrcrcrcrcr}
	J_{|\nu|}a_{l} & + & N_{|\nu|}b_{l} & + &
	J_{|\nu|}A_{l} & + & N_{|\nu|}B_{l} & = &
	(-i)^{|l|}J_{|l|} & + &
	\frac{f_{l}e^{ikr}}{\sqrt{r}} & + & \frac{g_{l}e^{ikr}}{\sqrt{r}}\\
	\pm J_{\pm(\nu+1)}a_{l} & \pm & N_{\pm(\nu+1)}b_{l} & \mp &
	J_{\pm(\nu+1)}A_{l} & \mp & N_{\pm(\nu+1)}B_{l} & = &
	\pm (-i)^{|l|}J_{\pm(l+1)} & - &
	\frac{if_{l}e^{ikr}}{\sqrt{r}} & + & \frac{ig_{l}e^{ikr}}{\sqrt{r}}\\
	\end{array}
\end{displaymath}
Where $+$($-$) corresponds to $\nu >$($<$)0 on the left-hand side, and
$l >$($<$)0 on the right-hand side.
Making use of the large $x$ form of the Bessel functions
\begin{eqnarray}
	J_{\mu}(x) \simeq  \sqrt{\frac{2}{\pi x}}
	\cos{(x-\frac{\mu\pi}{2}-\frac{\pi}{4})}, & &
	N_{\mu}(x) \simeq  \sqrt{\frac{2}{\pi x}}
	\sin{(x-\frac{\mu\pi}{2}-\frac{\pi}{4})},\\
	\nonumber
	\label{ping}
\end{eqnarray}
taking $0\geq \alpha_{R}>-1$ and eliminating $a_{l}$ and $A_{l}$ we find

\begin{eqnarray}
	f_{l} & = &
	\frac{1}{\sqrt{2\pi ki}}\left[-2ie^{\mp\frac{i\pi\nu}{2}}b_{l}
		-(e^{\mp i\pi l}-e^{\mp i\pi\nu})\right]
	\nonumber,\\
	 & & \nonumber\\
	g_{l} & = &
	\frac{1}{\sqrt{2\pi ki}}\left[-2ie^{\mp\frac{i\pi\nu}{2}}B_{l}\right]
	\nonumber,\\
	\nonumber
\end{eqnarray}
where the $-$($+$) now just corresponds to $\nu>$($<$)0.

\paragraph{(ii) -1\boldmath$<\nu<$0:}
In this instance, it makes the algebra easier if we use
$Z_{\nu}^{1}=J_{\nu}$ and
$Z_{\nu}^{2}=J_{-\nu}$. We cannot do this in the previous
case, however,  because if $\nu$ is an integer,
then $J_{\nu}$ and $J_{-\nu}$ are
linearly dependent. Since $-1<\nu<0$, then
$l=-1$ and $\alpha_{R}\neq 0$, so the choice is justified here.
Matching coefficients of $e^{il\phi}$ we
obtain
\begin{displaymath}
	\begin{array}{rcrcrcrcrcrcr}
	J_{\nu}a_{-1} & + & J_{-\nu}b_{-1} & + &
	J_{\nu}A_{-1} & + & J_{-\nu}B_{-1} & = &
	-iJ_{1} & + &
	\frac{f_{-1}e^{ikr}}{\sqrt{r}} & + & \frac{g_{-1}e^{ikr}}{\sqrt{r}}\\
	J_{\nu+1}a_{-1} & - & J_{-\nu-1}b_{-1} & - &
	J_{\nu+1}A_{-1} & + & J_{-\nu-1}B_{-1} & = &
	iJ_{0} & - &
	\frac{if_{-1}e^{ikr}}{\sqrt{r}} & + & \frac{ig_{-1}e^{ikr}}{\sqrt{r}}\\
	\end{array}
\end{displaymath}
and making use of large $x$ form of Bessel functions once more
we can now eliminate $a_{l}$ and $A_{l}$ to obtain $f_{l}$ and $g_{l}$
in terms of $b_{l}$ and $B_{l}$:
\begin{eqnarray}
	f_{l} & = &
	\frac{1}{\sqrt{2\pi ki}}\left[
	e^{-\frac{i\pi\nu}{2}}(e^{i\pi\nu}-e^{-i\pi\nu})b_{l} +
	(1+e^{-i\pi\nu})\right],\nonumber\\
	g_{l} & = &
	\frac{1}{\sqrt{2\pi ki}}\left[
	e^{-\frac{i\pi\nu}{2}}(e^{i\pi\nu}-e^{-i\pi\nu})B_{l}\right].
	\nonumber\\
	\nonumber
\end{eqnarray}

The asymptotic matching solutions also give us relationships between
$a_{l}$ and $b_{l}$, and, $A_{l}$ and $B_{l}$. Considering the
coefficients of $e^{-ikr}$ for the two cases we find that for
\begin{equation}
	\begin{array}{lclcl}
	\nu\geq0,-1\geq\nu: & &
	a_{l}+ib_{l} & = & (-i)^{\pm l}e^{\pm\frac{i\pi(l-\nu)}{2}},\\
	 &  & A_{l} + iB_{l} & = & 0,\\
 	 & & & &\\
	0>\nu>-1: & &
	e^{\frac{i\pi\nu}{2}}a_{l}+e^{-\frac{i\pi\nu}{2}}b_{l} & = &
	1,\\
	 & & e^{\frac{i\pi\nu}{2}}A_{l}
	+e^{-\frac{i\pi\nu}{2}}B_{l} & = & 0,\\
	\end{array}
\end{equation}
so we see that if either $a_{l}$ or $b_{l}$ dominates it will be of
order 1, whilst $A_{l}$ and $B_{l}$ are always identical up to a phase.

\subsection*{(d) Matching at $r=R$}

Now that we have solutions for inside and outside the core the obvious
next step is to match them at $r=R$ to give a solution valid for
all space.
Clearly all of $\chi_{1}^{l}$,$\chi_{2}^{l}$,$\psi_{1}^{l}$,$\psi_{2}^{l}$
will be continuous, though, due to the discontinuous
distribution of the string flux, this will not be the case for their
first derivatives.
{}From (\ref{eqs}), denoting the external (internal) solution by $+$
($-$), we see that

\begin{displaymath}
	\begin{array}{rcrcr}
	(\frac{d\chi_{1+}^{l}}{dr}-\frac{d\chi_{1-}^{l}}{dr}) & = &
	-\frac{\alpha_{R}}{R}\chi_{1}^{l} & + &
	m\psi_{2}^{l},\\
	 & & & & \\
	(\frac{d\chi_{2+}^{l}}{dr}-\frac{d\chi_{2-}^{l}}{dr}) & = &
	\frac{\alpha_{R}}{R}\chi_{2}^{l} & - &
	m\psi_{1}^{l},\\
	 & & & & \\
	(\frac{d\psi_{1+}^{l}}{dr}-\frac{d\psi_{1-}^{l}}{dr}) & = &
	-\frac{\alpha_{L}}{R}\psi_{1}^{l} & - &
	m\chi_{2}^{l},\\
	 & & & & \\
	(\frac{d\psi_{2+}^{l}}{dr}-\frac{d\psi_{2-}^{l}}{dr}) & = &
	\frac{\alpha_{L}}{R}\psi_{2}^{l} & + &
	m\chi_{1}^{l},\\
	\end{array}
\end{displaymath}
all evaluated at $r=R$. Substituting in our external and internal
solutions we find that the coefficients are related by
\begin{displaymath}
	\begin{array}{rcrcrcl}
	S^{+}_{1}a_{l} & + & S^{+}_{2}b_{l} & + &
	(\lambda_{l}S^{-}_{1}+S^{-}_{2})B_{l} & = & 0,\\
	T^{+}_{1}a_{l} & + & T^{+}_{2}b_{l} & - &
	(\lambda_{l}T^{-}_{1}+T^{-}_{2})B_{l} & = & 0,\\
	B^{+}U^{+}_{1}a_{l} & + & B^{+}U^{+}_{2}b_{l} & + &
	B^{-}(\lambda_{l}U^{-}_{1}+U^{-}_{2})B_{l} & = & 0,\\
	B^{+}V^{+}_{1}a_{l} & + & B^{+}V^{+}_{2}b_{l} & - &
	B^{-}(\lambda_{l}V^{-}_{1}+V^{-}_{2})B_{l} & = & 0,\\
	\end{array}
\end{displaymath}
where
\begin{displaymath}
	\begin{array}{rcrcc}
	S^{\pm}_{1,2} & = &
	Z^{1,2 '}_{\nu} & - &
	(\frac{J_{l}^{\prime}}{J_{l}}-\frac{\alpha_{R}}{R}+
	mB^{\pm}\frac{J_{l+2}}{J_{l+1}})Z^{1,2}_{\nu}\\
	\end{array}
\end{displaymath}
and we have similar expressions for $T$, $U$ and $V$ (the details are
contained in the appendix to this chapter).
We now have two cases to consider:

\paragraph{(i) \boldmath$m\ll k$:}

Since $\omega^{2}=m^{2}+k^{2}$ we see that
$B^{+}=\frac{\omega-k}{m}\simeq\frac{k}{m}(1+\frac{m^{2}}{2k^{2}})-
\frac{k}{m}\simeq0$ whilst $B^{-}\simeq 2$. Hence, the last two matching
solutions give that $B_{l}=0$, implying that in the zero mass limit
there is no helicity violation, as predicted earlier. We also find that
\begin{eqnarray}
	\frac{a_{l}}{b_{l}} & \simeq & -\frac{S^{+}_{2}}{S^{+}_{1}}.
\end{eqnarray}
It is straightforward to verify that
$S^{+}_{2}/S^{+}_{1}=T^{+}_{2}/T^{+}_{1}$, ensuring the
consistency of the first two matching solutions. By means of the
Bessel function relations (\ref{bess}) we see that
\begin{eqnarray}
	\frac{S^{+}_{2}}{S^{+}_{1}} & \simeq &
	\frac{\mp Z^{2}_{\pm(\nu+1)}J_{l}+Z^{2}_{\nu}J_{l+1}}
	{\mp Z^{1}_{\pm(\nu+1)}J_{l}+Z^{1}_{\nu}J_{l+1}}.
\end{eqnarray}
Making use, now, of the small argument form of Bessel functions, we find that
$b_{l}$ is suppressed by at least $(\omega R)^{2}$ with respect to
$a_{l}$ except for the case $0>\nu>-1$ corresponding to
$l=-1$, $\alpha_{R}\neq 0$ when the suppression is $(\omega
R)^{2(1+\nu)}$. Hence, it is a good approximation to ignore $b_{l}$ to
leading order.

\paragraph{(i) \boldmath$m\gg k$:} We now have $B^{+}\simeq B^{-}\simeq
1$ so, dropping the $\pm$, we can rewrite the first two matching
solutions as
\begin{equation}
	\begin{array}{rcrcr}
	S\left(\frac{b_{l}}{a_{l}}\right)
	& + & (\lambda_{l}+S)\left(\frac{B_{l}}{a_{l}}\right)
	& = & -1,\\
	 & & \\
	T\left(\frac{b_{l}}{a_{l}}\right)
	& - & (\lambda_{l}+T)\left(\frac{B_{l}}{a_{l}}\right)
	& = & -1,\\
	\end{array}
\end{equation}
where $S=S_{2}/S_{1}$ and $T=T_{2}/T_{1}$. From these
we see that if $T$ dominates $S$ then
\begin{equation}
	\begin{array}{ccccc}
	\frac{a_{l}}{b_{l}} & \simeq & \frac{a_{l}}{B_{l}} &
	\simeq & -2S,\\
	\end{array}
\end{equation}
whilst if $S$ dominates $T$
\begin{equation}
	\begin{array}{ccccc}
	\frac{a_{l}}{b_{l}} & \simeq & -\frac{a_{l}}{B_{l}} &
	\simeq & -2T.\\
	\end{array}
\end{equation}
Investigation of $S$ and $T$ reveals that for $\nu>-1$,  $T\gg S$,
whilst for $-1\leq \nu$, $S\gg T$. However, we need to know the
magnitude of $S$ and $T$ in order to determine the relative
suppression of $b_{l}$, and
$B_{l}$.
A summary of the results is
\begin{displaymath}
	\begin{array}{llll}
	T\gg S{\rm{:}} & \nu\geq 0{\rm{:}} &
	P\sim\frac{k^{2}}{\omega^{2}}(kR)^{-2(\nu+1)} &
	{\rm{Suppression\  of\  }}b_{l}{\rm{,\  }}B_{l}
	{\rm{\ greater\  than\  }}(\omega R)^{2}.\\
	 & & & \\
	 & 0>\nu >-1{\rm{:}} &
	P\sim (kR)^{-2\nu} &
	b_{l}{\rm{,\  }} B_{l}{\rm{\  dominate\  }}a_{l}.\\
	 & & & \\
	 & & & \\
	S\gg T{\rm{:}} & \nu=-1{\rm{:}} &
	S\simeq\frac{2}{\pi}\log{(kR)} &
	b_{l}{\rm{,\  }}B_{l}{\rm{\  relatively\ unsuppressed.}}\\
	 & & & \\
	 & -1>\nu{\rm{:}} &
	S\sim\frac{k^{2}}{\omega^{2}}(kR)^{2\nu} &
	{\rm{Suppression\  of\  }}b_{l}{\rm{,\  }}B_{l}
	{\rm{\ greater\  than\  }}(\omega R)^{2}.\\
	\end{array}
\end{displaymath}
We note that, except for when $l=-1$, we can ignore the contributions of
$b_{l}$ and $B_{l}$.
We now make use of this information to calculate the scattering
cross-sections.

\newpage

\section{The Scattering Cross-Sections}

The scattering amplitudes are given by the simple formulae
\begin{displaymath}
	\left.\frac{d\sigma}{d\phi}\right|_{-}^{+}  =
	\cases{{|f(\phi)|^{2}},&helicity conserving cross-section.\cr
	{}& \cr
	{|g(\phi)|^{2}},&helicity violating cross-section.}
\end{displaymath}
where $f(\phi)$ and $g(\phi)$ are given by
\begin{eqnarray}
	f(\phi)  =
	\sum_{l=-\infty}^{\infty}f_{l}e^{il\phi} & , &
	g(\phi)  =
	\sum_{l=-\infty}^{\infty}g_{l}e^{il\phi}.\nonumber\\
	\nonumber
\end{eqnarray}
Ignoring the effect of $b_{l}$ and $B_{l}$ for the time being, we find that
\begin{eqnarray}
	f(\phi) & = &
	-\frac{1}{\sqrt{2\pi ki}}\left[
	\sum_{l\geq 0} (1-e^{i\pi\alpha_{R}})e^{-il(\pi-\alpha_{R})}+
	\sum_{l\leq -1} (1-e^{-i\pi\alpha_{R}})e^{il(\pi+\alpha_{R})}
	\right]
	\nonumber\\
	 & = & \frac{ie^{-\frac{i\phi}{2}}}{\sqrt{2\pi ki}}
	\frac{\sin{(\pi\alpha_{R})}}{\cos{(\frac{\phi}{2})}},\nonumber\\
	g(\phi) & = & 0,\nonumber\\
	\nonumber
\end{eqnarray}
so the differential cross-sections for helicity
conservation and helicity flip are
\begin{eqnarray}
	\left.\frac{d\sigma}{d\phi}\right|_{+}  =
	\frac{1}{2\pi k}
	\frac{\sin^{2}{(\pi\alpha_{R})}}{\cos^{2}{(\frac{\phi}{2})}},
	& \hspace{1cm} &
	\left.\frac{d\sigma}{d\phi}\right|_{-}  =  0,\nonumber\\
	\nonumber
\end{eqnarray}
where we have denoted helicity conservation(flip) by $+$($-$). We
recognise the first as the full Aharonov-Bohm cross-section. We now
need to consider the effect of $b_{l}$ and $B_{l}$ for four
different cases. Note that in each one, however, it is only the $l=-1$
mode which will contribute, confirming the work by Ganoulis\cite{gan}.

\subsection*{(a) Non-integer $\alpha_{R}$}

\subsubsection*{(i) \boldmath$m\ll k$}

Since $B_{l}=0$ there is no helicity flip in this case, as
we predicted earlier by considering the zero mass limit of
$[H,\underline{\Sigma}.\underline{\Pi}]$.

The effect of the $b_{l}$ is to introduce a correction of order
$(kR)^{2(1+\alpha_{R})}$ such that
\begin{eqnarray}
	\left.\frac{d\sigma}{d\phi}\right|_{+} & = &
	\frac{1}{2\pi k}
	\frac{\sin^{2}{(\pi\alpha_{R})}}{\cos^{2}{(\frac{\phi}{2})}}
	(1+2{\rm{Re}}(\Delta))\\
	\nonumber
\end{eqnarray}
where
$\Delta  = 2ie^{i(\pi\alpha_{R}-\phi)/2}\cos{(\frac{\phi}{2})}b_{-1}$.

\subsubsection*{(ii) \boldmath$m\gg k$}

Here we have the curious case where $b_{l}$ and $B_{l}$ dominate for
the $l=-1$ mode, such that $b_{l}=B_{l}=e^{\frac{i\pi\nu}{2}}$. This
leads to
\begin{eqnarray}
	\left.\frac{d\sigma}{d\phi}\right|_{+}  =
	\frac{1}{2\pi k}
	\frac{\sin^{2}{(\pi\alpha_{R})}}{\cos^{2}{(\frac{\phi}{2})}}
	|2+e^{-i\phi}|^{2},
	 & \hspace{1cm} &
	\left.\frac{d\sigma}{d\phi}\right|_{-}  =
	\frac{2}{\pi k}\sin^{2}{(\pi\alpha_{R})},\nonumber\\
	\nonumber
\end{eqnarray}
so we have modified Aharonov-Bohm scattering, as predicted in \cite{gan}.

\subsection*{(b) Integer $\alpha_{R}$}

\subsubsection*{(i) \boldmath$m\ll k$}

Since $\alpha$ is an integer (in our case zero) we see that the
Aharonov-Bohm cross-section will vanish, and making use of our
matching solution, we see that, to {\cal{O}}$(kR)^2$, even the
correction due to $b_{l}$ vanishes. Hence, there appears to be no
scattering in this case. This is what we would expect, since if
$\alpha$ and the mass are zero, there is essentially no interaction
with the string, so we expect no scattering.

\subsubsection*{(ii) \boldmath$m\gg k$}

Once more, since $\alpha$ is an integer, the Aharonov-Bohm
cross-section will vanish. However, for $m\gg k$, $b_{l}$ and $B_{l}$
are relatively unsuppressed and we obtain Everetts cross-section for
both helicity conserving and helicity violating processes:
\begin{eqnarray}
	\left.\frac{d\sigma}{d\phi}\right|_{+} =
	\left.\frac{d\sigma}{d\phi}\right|_{-} & = &
	\frac{\pi}{8k}\frac{1}{[\log{(kR)}]^{2}}\\
	\nonumber
\end{eqnarray}

\vspace{15 mm}

\begin{tabular}{|l|l|l|}
	\multicolumn{3}{c}{\bf Summary of Results}\\
	\multicolumn{3}{c}{}\\ \hline
	$\alpha_{R}\notin Z$ & $m=0$ & Helicity conserving A-B
	cross-section.\\
	 & & Zero helicity flip cross-section.\\ \cline{2-3}
	 & $m\neq0$ &
	Modified A-B cross-section for both helicity conserving and
	flip processes.\\ \hline
	$\alpha_{R}\in Z$ & $m=0$ & No scattering.\\
	 & $m\neq0$ &
	Everett cross-section for both helicity conserving and
	flip processes.\\ \hline
\end{tabular}

\newpage

\section{Scattering from the Semi-local String}

As mentioned in the introduction, we obtain our
Lagrangian by modification of the Weinberg-Salaam model,
taking SU(2) charge and gauge fields to be zero. Working with quarks,
since they
have non-integer hypercharge, and so may display an Aharonov-Bohm
cross-section,
we use a Lagrangian with Yukawa type couplings
\begin{eqnarray}
	{\cal{L}} & = &
	i\bar{l}\gamma^{\mu}D_{\mu}^{L}l+
	i{\bar{u}}_{R}\gamma^{\mu}D_{\mu}^{R}u_{R}+
	i{\bar{d}}_{R}\gamma^{\mu}D_{\mu}^{R}d_{R}\nonumber\\
	 & &
	-f_{q}^{1}({\bar{u}}_{L},{\bar{d}}_{L})\tilde{\Phi}u_{R}+{\rm{h.c.}}
	\nonumber\\
	 & & -f_{q}^{2}({\bar{u}}_{L},{\bar{d}}_{L})\Phi d_{R}+{\rm{h.c.}}
	\nonumber\\
	\nonumber
\end{eqnarray}
where $\tilde{\Phi}=i\tau_{2}\Phi$ and
we assume that $f_{q}^{1,2}$ are real.
It is seen that this is invariant under
\begin{displaymath}
	\begin{array}{cll}
	\left.
	\begin{array}{c}
	e_{R}\\
	e_{L}\\
	\Phi\\
	\end{array}
	\right\}
	\rightarrow U
	\left\{
	\begin{array}{c}
	e_{R}\\
	e_{L}\\
	\Phi\\
	\end{array}
	\right. &
	{\rm{where\  }}U=e^{i\beta(x)Y}\in U(1); &
	\rm{local\  U(1)\ symmetry.}\\
 	 & & \\
	\left.
	\begin{array}{c}
	e_{L}\\
	\Phi\\
	\end{array}
	\right\}
	\rightarrow S
	\left\{
	\begin{array}{c}
	e_{L}\\
	\Phi\\
	\end{array}
	\right. &
	{\rm{where\  }}S\in SU(2); &
	\rm{global\  SU(2)\ symmetry.}\\
	\end{array}
\end{displaymath}
Note that hypercharge is conserved.
Such a model has a string solution identical to that in the
electroweak case, the only difference being that here we write $B$
instead of $Z$ for the relevant gauge field. This comes as no surprise
as in both cases we are embedding the same Nielsen-Olesen vortex in a
``larger'' model.

Considering the case
of the down-quark, a little algebra
yields the equations of motion
\begin{displaymath}
	\begin{array}{rcrcr}
	i\gamma^{\mu}D_{\mu}^{L}d_{L} & - &
	f_{q}^{2}fd_{R}e^{i\phi} & = & 0\\
	i\gamma^{\mu}D_{\mu}^{R}d_{R} & - &
	f_{q}^{2}fd_{L}e^{-i\phi} & = & 0\\
	\end{array}
\end{displaymath}
which if we use the usual Dirac representation for the gamma matrices,
write $d_{L}=(0,\psi)$,$d_{R}=(\chi,0)$, restrict motion to the plane
perpendicular to the z-axis, and set $\partial_{t}=-i\omega$ become

\begin{equation}
	\begin{array}{lclclcc}
 	\omega \chi & + & i\sigma^{j} D_{j}^{R}\chi & - &
		f_{e}fe^{-i\phi}\psi & = & 0 \\
	\omega \psi & - & i\sigma^{j} D_{j}^{L}\psi & - &
		f_{e}fe^{i\phi}\chi & = & 0 \\
	\end{array}
	\label{sl}
\end{equation}
These are of the same form as (\ref{ganeq}) but with $\alpha_{L(R)}$ replaced
by
$Y_{L(R)}$. Since, in addition, $Y_{L}=Y_{R}+1$, we can make
use of the results previously
calculated for the electroweak string.

For the down quark $Y_{d_{R}}=-\frac{2}{3}\notin Z$, so
the dominant scattering is Aharonov-Bohm, and the cross-sections are
identical to those for the electroweak string in the case of
non-integer flux. Since our calculations were done assuming that
$0\geq\alpha_{R}>-1$ we can not immediately apply our results to the
case of the electron, which has hypercharge $Y_{e_{R}}=-2$. However, we
see that we should be able to modify our calculation by
shunting the partial waves
two along without physically altering the solution. Hence, we would
expect the electron to display an Everett cross-section. The case of
the up-quark is a little more tricky since we no longer have the
relation $Y_{L}=Y_{R}+1$ (in fact $Y_{L}=Y_{R}-1$
here).
This is not easily incorporated into our solution, and the most we can
say is that, physically, we would expect the up quark to behave in a
similar manner to the down-quark, since they both possess non-integer
hypercharge, and display an Aharonov-Bohm interaction.

\newpage

\section{Conclusions}

We have investigated elastic scattering off embedded
defects like electroweak and semi-local strings, and, using the `top-hat'
model of the core, have been able to obtain a fuller solution to the
scattering problem than previously found, whilst confirming existing results.

In the massive case, for fractional string flux, we have found a
modified Aharonov-Bohm cross-section for both the
helicity conserving and helicity-flip processes, whilst,
for integer flux we have obtained the
same cross-section as that of Everett \cite{eve}, ie a strong interaction
cross-section, but with logarithmic suppression. This latter case may be
important for electron scattering from semi-local strings, whlist our
techniques and results are also applicable to grand unified
strings and chiral fermions. In the limit as the mass goes to zero, we
see that helicity violating processes disappear, suggesting that such
effects may be stronger at lower energies.

We stress that for fractional flux the total cross-section is a strong
interaction cross-section, independent of the core radius. For integer
flux the core radius appears in the logarithmic suppression factor,
but this is only a
mild dependence. Hence the evolution of these defects will be
the same as that of local strings, with the strings strongly interacting with
the surrounding plasma during the friction dominated era.

The significance of our result is that it clarifies
the concept of Aharonov-Bohm phase for a particle whose
internal degrees of freedom couple to a flux tube, or string
(in our example in a
left-right asymmetric way). Hence, if cosmic strings are ever found it
will be possible to study their interactions with matter and maybe even
``probe'' their core through this type of scattering experiment. Our
results may also have implications for grand unified strings. This is
currently under investigation \cite{rach}.

One of us (A.P.M.) acknowledges S.E.R.C. for financial support.

\newpage

\section*{Appendix: Matching at \boldmath{$r=R$}}

Our solution for $r<R$ is

\begin{displaymath}
	\left(
	\begin{array}{c}
	c_{l}J_{l}e^{il\phi}(\omega r)\\
	ic_{l}J_{l+1}e^{i(l+1)\phi}(\omega r)\\
	d_{l}J_{l+1}e^{i(l+1)\phi}(\omega r)\\
	id_{l}J_{l+2}e^{i(l+2)\phi}(\omega r)\\
	\end{array}
	\right)
\end{displaymath}
whilst our  external solution is
\begin{displaymath}
	\left(
	\begin{array}{ccccccc}
	(a_{l}Z_{\nu}^{1}(k R) & + &
	b_{l}Z_{\nu}^{2}(k R) & + &
	A_{l}Z_{\nu}^{1}(k R) & + &
	B_{l}Z_{\nu}^{2}(k R))e^{il\phi}\\
	i(a_{l}Z_{(\nu+1)}^{1}(k R) & + &
	b_{l}Z_{(\nu+1)}^{2}(k R) & - &
	A_{l}Z_{(\nu+1)}^{1}(k R) & - &
	B_{l}Z_{(\nu+1)}^{2}(k R))e^{i(l+1)\phi}\\
	(a_{l}B^{+}Z_{\nu}^{1}(k R) & + &
	b_{l}B^{+}Z_{\nu}^{2}(k R) & + &
	A_{l}B^{-}Z_{\nu}^{1}(k R) & + &
	B_{l}B^{-}Z_{\nu}^{2}(k R))e^{i(l+1)\phi}\\
	i(a_{l}B^{+}Z_{(\nu+1)}^{1}(k R) & + &
	b_{l}B^{+}Z_{(\nu+1)}^{2}(k R) & - &
	A_{l}B^{-}Z_{(\nu+1)}^{1}(k R) & - &
	B_{l}B^{-}Z_{(\nu+1)}^{2}(k R))e^{i(l+2)\phi}\\
	\end{array}
	\right)
	\nonumber
	\label{doog}
\end{displaymath}
so by continuity
\begin{displaymath}
	\begin{array}{rcrcrcrcr}
	Z_{\nu}^{1}a_{l} & + &
	Z_{\nu}^{2}b_{l} & + &
	Z_{\nu}^{1}A_{l} & + &
	Z_{\nu}^{2}B_{l} & = &
	J_{l}c_{l} \\
	Z_{(\nu+1)}^{1}a_{l} & + &
	Z_{(\nu+1)}^{2}b_{l} & - &
	Z_{(\nu+1)}^{1}A_{l} & - &
	Z_{(\nu+1)}^{2}B_{l} & = &
	J_{l+1}c_{l} \\
	B^{+}Z_{\nu}^{1}a_{l} & + &
	B^{+}Z_{\nu}^{2}b_{l} & + &
	B^{-}Z_{\nu}^{1}A_{l} & + &
	B^{-}Z_{\nu}^{2}B_{l} & = &
	J_{l+1}d_{l}\\
	B^{+}Z_{(\nu+1)}^{1}a_{l} & + &
	B^{+}Z_{(\nu+1)}^{2}b_{l} & - &
	B^{-}Z_{(\nu+1)}^{1}A_{l} & - &
	B^{-}Z_{(\nu+1)}^{2}B_{l} & = &
	J_{l+2}d_{l}\\
	\end{array}
\end{displaymath}
whilst the first derivatives give us
\begin{displaymath}
	\begin{array}{r}
	Z_{\nu}^{1'}a_{l}  + \hspace{12 mm}
	Z_{\nu}^{2'}b_{l}
	+ \hspace{10 mm}
	Z_{\nu}^{1'}A_{l}  + \hspace{11 mm}
	Z_{\nu}^{2'}B_{l} \hspace{2mm} = \hspace{6 mm}
	(J_{l}^{'}-\frac{\alpha_{R}}{R}J_{l})c_{l}  +\hspace{17 mm}
	mJ_{l+2}d_{l}\\
	Z_{(\nu+1)}^{1'}a_{l}  + \hspace{6 mm}
	Z_{(\nu+1)}^{2'}b_{l}
	 - \hspace{6 mm}
	Z_{(\nu+1)}^{1'}A_{l}  - \hspace{5 mm}
	Z_{(\nu+1)}^{2'}B_{l} \hspace{2mm} =
	(J_{l+1}^{'}+\frac{\alpha_{R}}{R}J_{l+1})c_{l}  - \hspace{16 mm}
	mJ_{l+1}d_{l}\\
	B^{+}Z_{\nu}^{1'}a_{l}  + \hspace{5 mm}
	B^{+}Z_{\nu}^{2'}b_{l}
	- \hspace{5 mm}
	B^{-}Z_{\nu}^{1'}A_{l}  + \hspace{5 mm}
	B^{-}Z_{\nu}^{2'}B_{l} \hspace{2mm} = \hspace{13 mm}
	-mJ_{l+1}c_{l}  +
	(J_{l+1}^{'}-\frac{\alpha_{L}}{R}J_{l+1})d_{l}\\
	B^{+}Z_{(\nu+1)}^{1'}a_{l}  +
	B^{+}Z_{(\nu+1)}^{2'}b_{l}
	-
	B^{-}Z_{(\nu+1)}^{1'}A_{l}  -
	B^{-}Z_{(\nu+1)}^{2'}B_{l} \hspace{2mm} =
	\hspace{20 mm}mJ_{l}c_{l}  +
	(J_{l+2}^{'}+\frac{\alpha_{L}}{R}J_{l+2})d_{l}\\
	\end{array}
\end{displaymath}
Note that here $^{\prime}$ denotes $\frac{d}{dr}$. We now make use of
the fact that $1/B^{+}=B^{-}$ to rewrite our
matching conditions as
\begin{displaymath}
	\begin{array}{rcrcrcl}
	S^{+}_{1}a_{l} & + & S^{+}_{2}b_{l} & + &
	(\lambda_{l}S^{-}_{1}+S^{-}_{2})B_{l} & = & 0\\
	T^{+}_{1}a_{l} & + & T^{+}_{2}b_{l} & - &
	(\lambda_{l}T^{-}_{1}+T^{-}_{2})B_{l} & = & 0\\
	B^{+}U^{+}_{1}a_{l} & + & B^{+}U^{+}_{2}b_{l} & + &
	B^{-}(\lambda_{l}U^{-}_{1}+U^{-}_{2})B_{l} & = & 0\\
	B^{+}V^{+}_{1}a_{l} & + & B^{+}V^{+}_{2}b_{l} & - &
	B^{-}(\lambda_{l}V^{-}_{1}+V^{-}_{2})B_{l} & = & 0\\
	\end{array}
\end{displaymath}
where
\begin{displaymath}
	\begin{array}{rcrcc}
	S^{\pm}_{1,2} & = &
	Z^{1,2 '}_{\nu} & - &
	(\frac{J_{l}^{\prime}}{J_{l}}-\frac{\alpha_{R}}{R}+
	mB^{\pm}\frac{J_{l+2}}{J_{l+1}})Z^{1,2}_{\nu}\\
	T^{\pm}_{1,2} & = &
	Z^{1,2 '}_{(\nu+1)} & - &
	(\frac{J_{l+1}^{\prime}}{J_{l+1}}+\frac{\alpha_{R}}{R}-
	mB^{\pm}\frac{J_{l+1}}{J_{l+2}})Z^{1,2}_{(\nu+1)}\\
	U^{\pm}_{1,2} & = &
	Z^{1,2 '}_{\nu} & - &
	(\frac{J_{l+1}^{\prime}}{J_{l+1}}-\frac{\alpha_{L}}{R}-
	mB^{\mp}\frac{J_{l+1}}{J_{l}})Z^{1,2}_{\nu}\\
	V^{\pm}_{1,2} & = &
	Z^{1,2 '}_{(\nu+1)} & - &
	(\frac{J_{l+2}^{\prime}}{J_{l+2}}+\frac{\alpha_{L}}{R}+
	mB^{\mp}\frac{J_{l+l}}{J_{l+1}})Z^{1,2}_{(\nu+1)}\\
	\end{array}
\end{displaymath}

\end{document}